\documentclass[11pt]{article}
\usepackage{mathdots}
\usepackage{color,graphicx}
\usepackage{amssymb,amsmath}
\usepackage{mathrsfs}
\usepackage{wrapfig,framed,cancel}
\usepackage[colorlinks=true, pdfstartview=FitV, linkcolor=black, citecolor=black, urlcolor=blue]{hyperref}
\definecolor{shadecolor}{rgb}{0.93, 0.93, 0.86}
\definecolor{light-blue}{rgb}{0.8,0.85,1}
\definecolor{blue}{rgb}{0,0,1}
\definecolor{red}{rgb}{1,0,0}

\allowdisplaybreaks \numberwithin{equation}{section}
\newtheorem{thm}{Theorem}[section]
\newtheorem{prp}[thm]{Proposition}
\newtheorem{lem}[thm]{Lemma}
\newtheorem{dfn}[thm]{Definition}
\newenvironment{defn}{\begin{dfn} \rm }{\end{dfn}}
\newtheorem{cor}[thm]{Corollary}
\newtheorem{example}[thm]{Example}
\newenvironment{exa}{\begin{example} \rm }{ \end{example}}
\newtheorem{remark}[thm]{Remark}

\newtheorem{problem}{Problem}[section]
\newenvironment{rmk}{\begin{remark} \rm }{\hfill $\Box$ \end{remark}}
\newenvironment{prf}{\noindent {\it Proof:} \ }{\hfill $\Box$}
\newenvironment{prfof}[1]{\noindent {\it Proof of #1} \ }{\hfill $\Box$}

\newcommand\od{\mathrm{d}}
\newcommand\ad{\mathrm{ad}}

\newcommand{\nn}{\nonumber}
\newcommand\pd{\partial}
 \newcommand{\Ld}{\Lambda}
\newcommand{\al}{\alpha}
\newcommand{\gm}{\gamma}

\newcommand{\om}{\omega} 
\newcommand{\Gm}{\Gamma}

\newcommand{\dt}{\delta}
 \newcommand{\Ta}{\Theta}

\newcommand{\res}{\mathrm{res}}

\newcommand{\im}{\mathrm{Im}}

\newcommand\sL{\mathscr{L}} 
\newcommand\fg{\mathfrak{g}} \newcommand\cH{\mathcal{H}}
\newcommand\Z{\mathbb{Z}}
\newcommand\C{\mathbb{C}}
\newcommand\R{\mathbb{R}}
\newcommand\N{\mathbb{N}}
\newcommand\bP{\mathbb{P}}

\newcommand{\set}[1]{\left\{#1\right\}}
\newcommand\ra{\rangle}
\newcommand\la{\langle}

\newcommand{\p}{\partial}

\newcommand{\bt}{\mathbf{t}}  
\newcommand{\rs}{\mathrm{s}}

\def\Gr{\mathrm{Gr}}
\def\Id{\mathrm{Id}}

\def\bea{\begin{eqnarray}}
\def\eea{\end{eqnarray}}

\begin{document}

\title{Tau functions and the limit of block Toeplitz determinants}
\author{Mattia Cafasso$^{\dagger}$ and  Chao-Zhong Wu$^{\ddagger}$ \\
   \footnotesize $\dagger$ \footnotesize LUNAM Universit\'e, LAREMA, Universit\'e d'Angers\\
    \footnotesize 2 Boulevard Lavoisier, 49045 Angers, France. cafasso@math.univ-angers.fr \\
   $\ddagger$
   \footnotesize School of Mathematics and Computational Science, Sun Yat-sen University \\
   \footnotesize Guangzhou 510275, P.R. China.
 wuchaozhong@sysu.edu.cn }

\date{}
\maketitle

\begin{abstract}
A classical way to introduce tau functions for integrable
hierarchies of solitonic equations is by means of the
Sato--Segal--Wilson infinite-dimensional Grassmannian. Every point
in the Grassmannian is naturally related to a Riemann--Hilbert
problem on the unit circle, for which Bertola proposed a tau
function that generalizes the Jimbo--Miwa--Ueno tau function for
isomonodromic deformation problems. In this paper, we prove that the
Sato--Segal--Wilson tau function and the (generalized)
Jimbo--Miwa--Ueno isomonodromy tau function coincide under a very
general setting, by identifying each of them to the large-size limit
of a block Toeplitz determinant. As an application, we give a new
definition of tau function for Drinfeld--Sokolov hierarchies (and
their generalizations) by means of infinite-dimensional
Grassmannians, and clarify their relation with other % definitions of
tau functions given in the literature.
\end{abstract}

\tableofcontents

\section{Introduction}

Tau functions were introduced in 1980s by the Kyoto school in the
study of integrable equations. Since the very beginning (see for
instance \cite{JMMS}), tau functions have been applied to other
branches of mathematical physics, especially in statistical physics
and, more recently, in the theory of random matrices and
determinantal point processes. A classical way to introduce tau
functions for integrable hierarchies of solitonic equations is by
means of the Sato--Segal--Wilson infinite-dimensional Grassmannian
\cite{Sa,SW} whose points are sub-spaces of $L^2(S^1)$. On the other
hand, it is known that a point in the Sato--Segal--Wilson
Grassmannian is naturally related to a Riemann--Hilbert problem on
the unit circle (see for instance \cite{Sat} or the more recent
\cite{Ca}), and to an arbitrary (sufficiently regular)
Riemann--Hilbert problem, Bertola \cite{Be} proposed a tau function
that generalizes the Jimbo--Miwa--Ueno tau function \cite{JMU} for
isomonodromic deformation problems. Hence, a natural question is:
what is the relationship between the Sato--Segal--Wilson tau
function and the (generalized) Jimbo--Miwa--Ueno isomonodromy tau
function?

In this article, we will show (see Theorem \ref{main} below) that
the two tau functions above coincide under a very general setting,
which includes not only the case of Gelfand--Dickey hierarchies but
also \emph{arbitrary} Drinfeld--Sokolov hierarchies \cite{DS} and
some more recent generalizations of them \cite{GHM, Mi, FHM}. As an
immediate byproduct (Corollary \ref{thm-tauw}), we will obtain, in
this general setting, a formula that generalizes the so-called Sato
formula (see \eqref{wKP} for the case of the Kadomtsev--Petviashvili
(KP) hierarchy). This formula connects the matrix Baker function and
the Sato--Segal--Wilson tau function for the Grassmannian.

The proof of our main result is divided into two parts. Namely, we
will identify each of the Sato--Segal--Wilson and the
Jimbo--Miwa--Ueno tau function with the large-size limit of a block
Toeplitz determinant \cite{Wi74,Wi75,Wi76}. The first part (the
equivalence between the Sato--Segal--Wilson tau function and the
limit of a block Toeplitz determinant) is a natural generalization
of the work by one of the author \cite{Ca}, who proved the
equivalence in the case of Gelfand--Dickey (or $n$-reduced KP)
hierarchies. Here we will use a different method, and the proof is
much simplified. The second part is based on a theorem by Widom
\cite{Wi76} (rederived by Its, Jin and Korepin in \cite{IJK}) that
links the Szeg\"o--Widom asymptotic formula with the
Riemann--Hilbert factorization of the related symbol, see
Theorems~\ref{thm-SW} and~\ref{factorization} below.

As mentioned above, the Gelfand--Dickey hierarchies are contained in
a more general class of integrable hierarchies proposed by Drinfeld
and Sokolov \cite{DS}. In fact, Drinfeld and Sokolov constructed a
Hamiltonian integrable hierarchy for every affine Kac--Moody
algebra, and recovered the Gelfand--Dickey  hierarchies when the the
affine algebra is of type $A_n^{(1)}$. This construction was later
generalized by de Groot, Hollowood and Miramontes \cite{GHM, Mi}
(also by Feher, Harnad and Marshall \cite{FHM}), who replaced the
principal Heisenberg subalgebra by an arbitrary Heisenberg
subalgebra corresponding to a certain gradation \cite{KP} of the
affine algebra. Tau functions for the Drinfeld--Sokolov hierarchies
as well as their generalizations were investigated from several
different points of view, e.g. representation theory \cite{HM},
Hamiltonian formalism \cite{Mi, LWZ, Wu} and algebraic geometry
\cite{BF, Sa}. However, a Sato--Segal--Wilson construction for
arbitrary Drinfeld--Sokolov hierarchies (and generalizations) is
still missing, though some results in this direction can be found in
\cite{Mi}. We will deal with this issue here. What is more, we will
clarify the relationship between the Sato--Segal--Wilson tau
function and the tau function of Drinfeld--Sokolov hierarchies
defined by one of the authors in \cite{Wu}. Hence, via the results
in \cite{Wu, Sa}, one can see the relation between the
Sato--Segal--Wilson tau function and the tau functions introduced in
the literature \cite{EF, HM, Mi, Sa} from different background.

This article is organized as follows. In the following section, we
will first recall the definition of the Sato-Segal-Wilson
Grassmannian, as well as the Baker function and tau function
associated to it. Secondly, the related Riemann--Hilbert problem and
its generalized Jimbo-Miwa-Ueno isomonodromy tau function will be
considered; our main result will also be formulated. The proof of
the main result will be given in Section~3, based on the properties
of large-size block Toeplitz determinant. In Section~4, we will
recall the tau function defined in \cite{Wu} for Drinfeld--Sokolov
hierarchies associated to affine Kac--Moody algebras, then give a
new definition of their tau functions via Sato-Segal-Wilson
Grassmannian, which will be extended to the case of generalized
Drinfeld--Sokolov hierarchies.

\section{Grassmannians and Riemann--Hilbert problems}\label{sec-SSW}

In this section we will review the notions of Baker function and tau
function associated to the Sato--Segal--Wilson Grassmannian,
and consider a related Riemann-Hilbert problem.

\subsection{The Sato--Segal--Wilson Grassmannian}

Let us recall some basic facts and notations about the
Sato--Segal--Wilson infinite-dimensional Grassmannian, following
the works \cite{SW, PS, Dickey}. We start with the Hilbert space of
vector-valued $L^2$ functions
\begin{equation}\label{}
H^{(n)}:=L^2(S^1,\C^n)=H^{(n)}_+\oplus H^{(n)}_-
\end{equation}
 where a function $v \in H^{(n)}$ is defined through its Fourier expansion $v(z)=\sum_k
v_k z^k$ with $v_k$ being column vectors, and $H^{(n)}_+$ (resp.
$H^{(n)}_-$) is the subspace of series with non-negative (resp.
negative) Fourier coefficients. Note, in particular, that elements
in $H_+^{(n)}$ are boundary values of holomorphic functions on the
disc $D_0 := \{z \in \C\mid |z| < 1\}$, while elements in
$H_-^{(n)}$ are boundary values of holomorphic functions on the disc
$D_\infty := \{z \in \bP_\C \mid |z| > 1\}$.
%\textcolor{blue}{Chaozhong here the
%reason is the following: take an element $v \in H_+^{(n)}$; at least
%there is a point $z_0 \in S^1$ in which the series defining $v$
%converges; but then, for analytical continuation, it converges for
%any $z$ such that $|z| < |z_0|$, hence it defines an analytical
%function on $D_0$. For elements in $H^{(n)}_-$, you repeat the same
%proof substituting $z$ with $1/z$. Since it is written on \cite{SW}
%(page 6), I think we can safely state it without explanation. Shame
%on me, I did not realize it before!}
We also denote with $p_\pm$ the
projections of $H^{(n)}$ onto its two subspaces $H^{(n)}_\pm$.
%\textcolor{blue}{Chaozhong I slightly shortened your version, I
%think that persons will understand also with shorter statement...}.

Clearly the vector space $H^{(n)}$ is spanned by the standard
vectors $\{z^k{\rm e}_\al\mid \al=1,\ldots,n; k\in\Z\}$, where ${\rm
e_\al}$ is the column vector with its $\al$-th component being $1$
and the other components vanish. In other words, we can fix a basis
of $H^{(n)}=H^{(n)}_-\oplus H^{(n)}_+$ whose elements are the
columns of the following matrix in block form:
\begin{equation}\label{basis}
\left( \dots, z^{-2}I_n, z^{-1}I_n, \ I_n, z I_n, z^2 I_n,
\dots\right),
\end{equation}
where $I_n=({\rm e}_1, \dots, {\rm e}_n)$ is the unit matrix. The
bases of the subspaces $H^{(n)}_-$ and $H^{(n)}_+$ are fixed
accordingly.

With respect to the basis given by \eqref{basis}, any vector
$v(z)=\sum_k v_k z^k$ in $H^{(n)}$ can be identified with its
coordinates as
\begin{equation}\label{convention}
    v \sim \left(\begin{array}{c}\vdots \\ v_{-1} \\ v_0 \\ v_1 \\ \vdots
    \end{array}\right).
\end{equation}
Observe that our convention, for the purpose of comparing with
Toeplitz matrices below, is slightly different from the one used in
\cite{SW}, where the index of Fourier coefficients increases going
upwards.

\begin{dfn}[ \cite{SW}]
    The Sato--Segal--Wilson Grassmannian $\Gr^{(n)}$ is the collection of subspaces\\
    $W\subseteq H^{(n)}$ such that
    \begin{enumerate}
        \item the projection $p_+: W\to H^{(n)}_+$ is a Fredholm operator,
        namely, both its kernel and its cokernel are of finite dimension;
        \item the projection $p_-:W\to H^{(n)}_-$ is a compact
        operator;
        \item $z W\subseteq W$.
    \end{enumerate}
\end{dfn}

The Grassmannian $\Gr^{(n)}$ has a deep relationship with loop
groups \cite{PS, SW}. Let
\[
LGL_n := \{\gamma: S^1\to GL_n\}
\]
 be the group of invertible
continuous  loops. It acts naturally on $H$ by multiplication as
\begin{eqnarray}
LGL_n\times H^{(n)}& \longrightarrow& H^{(n)}
    \nonumber\\
    (\gamma,v)&\longmapsto & \gamma\cdot v. \label{GH}
\end{eqnarray}
According to \cite{PS}, one can modify $LGL_n$ to a group of some
particular measurable loops and makes the action \eqref{GH}
transitive on $\Gr^{(n)}$. More precisely, given a measurable loop $\gamma(z) = \sum_k \gamma_k z^k$, introduce the following two norms:
\begin{equation}
\|\gamma\|_\infty:=\mathrm{ess}\sup_{|z|=1}\|\gamma(z)\|\quad\quad
\|\gamma\|_{2,1/2}:=\sum_k\Big(\left|k\right|\|\gamma_k\|^2\Big)^{1/2},
\end{equation}
where we denoted with $\|\cdot\|$ the standard Hilbert--Schmidt norm.

\begin{defn}
The loop group $L_{1/2}GL_n$ is composed of all invertible
measurable loops $\gamma$ taking value in $GL_n$ such that
\[
 \|\gamma\|_\infty+\|\gamma\|_{2,1/2}<\infty.
 \]

 Analogously, we denote with $L_{1/2}U_n$ its subgroup consisting of loops taking values in the unitary group.
\end{defn}
%The following is \cite{PS}:
\begin{thm}[Theorem 8.3.2 in \cite{PS}]\label{thm-Un}
    The group $L_{1/2}U_n$ acts transitively on $\Gr^{(n)}$, and the isotropy group
    of $H_+^{(n)}$ is $U_n$, i.e., the group of constant loops.
\end{thm}

A subspace $W\in\Gr^{(n)}$ is said to be transversal (to
$H^{(n)}_-$) if
\[
\left.p_+\right|_{W}: W\longrightarrow H^{(n)}_+
\]
is a one-to-one correspondence. All transversal subspaces in
$\Gr^{(n)}$ compose the so-called big cell of the Grassmannian,
which will be denoted as $\Gr^0_{(n)} = \Gr^0$, following the
notation in \cite{KS} and omitting the subscript ``$(n)$'' when no
ambiguity arises. In particular, we can associate to every element
of the big cell $\Gr^0$ a loop with some definite asymptotic
property.
\begin{cor}\label{cor1}
    For any $W \in \Gr^0_{(n)}$, it exists a unique loop $\gamma \in L_{1/2}U_n$
    such that $$W = \gamma H^{(n)}_+;\quad p_+(\gamma) = \Id.$$
\end{cor}
\begin{prf}
According to Theorem~\ref{thm-Un}, it exists $\tilde\gamma \in
L_{1/2}U_n$ such that
    $\tilde\gamma H_+^{(n)} = W$. Since $W \in \Gr^{0}$,
we have $p_+(\tilde\gamma) = \tilde\gamma_0$, with $\tilde\gamma_0$
being constant and invertible (Otherwise %Suppose in fact that this is not the case, then
there would not be $n$ linear independent vectors $\{w_\al\mid \al
=1,\ldots,n\}$ in $W$ such that $p_+(w_\al) ={\rm e}_\al$ for $\al =
1,\ldots,n$). Now we take $\gamma(z):= \tilde\gamma(z)
\tilde\gamma_0^{-1}$, then we obtain  $\gamma$ with the right
asymptotic property and, of course, such that $\gamma H^{(n)}_+ =
W$.
%, since $\gamma_0^{-1}$ acts trivially on $H^{(n)}_+$.

For the uniqueness, suppose that there exist
$\gamma,\tilde{\gamma}\in L_{1/2}U_n$ satisfying
$p_+(\gamma)=p_+(\tilde{\gamma})=\Id$ and $\gamma H^{(n)}_+ =
\tilde{\gamma} H^{(n)}_+ = W$. Let  $\gamma^{[\al]}$ and
$\tilde{\gamma}^{[\al]}$ denote the $\al$-th columns of these loops.
Clearly, $\gamma^{[\al]} - \tilde{\gamma}^{[\al]} \in W$. Since $p_+
(\gamma^{[\al]} - \tilde{\gamma}^{[\al]}) = 0$, and $p_+|_W$ is
injective, then $\gamma^{[\al]} = \tilde{\gamma}^{[\al]}$. Note that
$\al$ is arbitrary, thus we obtain $\gamma= \tilde{\gamma}$.
 \end{prf}

Now denote with $G_+$ the group of loops $g(z)$ in $L_{1/2}U_n$
which extend analytically to the whole plane such that $g(0)\in
U_n$, and consider an (arbitrary) abelian subgroup $G^a_+ \subseteq
G_+$. The following definition is taken from \cite{Dickey} and it is
a generalization of the one in \cite{SW} for the matrix case. There,
we say that a matrix-valued function belongs to a certain point $W
\in \Gr^{0}_{(n)}$ if all the columns do.

\begin{dfn}\label{def-bf}
Suppose an element $W\in \Gr^0_{(n)}$ is given. A matrix function
$w(g;z)$, depending on $g\in G_+^a$ and $z\in S^1$, is called the
Baker function associated to $W$ if
\begin{enumerate}
        \item $w(g;z)\in W$ for almost all $g\in G_+^a$;
        \item $p_+(g^{-1}w(g;z)) = \Id$.
\end{enumerate}
\end{dfn}
The definition above needs some clarifications. When we write that
``\emph{$w(g;z)\in W$ for almost all $g\in G_+^a$}'' we mean that
this is true for every $g \in G_+^{a}$ such that $g^{-1}W$ is
transversal to $H_-^{(n)}$ (which is a condition generically
satisfied, see \cite{SW}). Then the existence of $w(g;z)$, when
$g^{-1}W$ is transversal to $H^{(n)}_-$, is deduced from Corollary
\ref{cor1}. Indeed, let $\varphi_g \in L_{1/2}U_n$ such that
$g^{-1}W = \varphi_g H^{(n)}_+$ and $p_+(\varphi_g) = \Id$. We just
define $w(g;z) := g(z)\varphi_g(z)$, and it is immediately seen that
the two conditions above are satisfied. Note also that, from the
unicity of $\varphi_g$, we deduce the uniqueness of $w(g;z)$, and
this explains why we speak about \emph{the} Baker function
associated to $W$. We remark that the uniqueness of the Baker function
can also be derived from the uniqueness of solution of the
Riemann--Hilbert problem in Proposition~\ref{thm-bfRH} below.

%\begin{remark}
%The discussion above gives a relation between Baker functions and Riemann--Hilbert boundary value problems; suppose in fact that it exists $\varphi_g$ such that
%$$g^{-1}W = g^{-1}\gamma H^{(n)}_+ = \varphi_g H^{(n)}_+.$$
%This means, in particular, that it exists an other loop $\psi_g$ such that
%\begin{equation}\label{preliminaryRH}
%   \varphi_g  = (g^{-1}\gamma)\psi_g; \quad p_+(\varphi_g) = \Id, \quad p_-(\psi_g) = 0.
%\end{equation}
%Indeed, for every $i = 1,...,n$, the i-th column $\varphi_g^{[i]}$ belongs to $g^{-1}\gamma H^{(n)}_+$, hence it is of the form $\varphi_g^{[i]} = (g^{-1}\gamma) v_i$; with $v_i \in H^{(n)}_+$.
%On the other hand equations \eqref{preliminaryRH}, in the case where all the terms are sectionally-analytic functions,  are the typical ones defining a Riemann--Hilbert boundary value problem on the circle. On the next subsection we will elaborate on this relation.
%\end{remark}

Now we proceed to recall the definition of the Sato--Segal--Wilson
tau function for a point in the Grassmannian $\Gr^0_{(n)}$ acted by
an abelian group $G_+^a$. For convenience, we will denote with
$W_\gamma$ the point corresponding to the (unique) $\gamma \in
L_{1/2}U_n$ such that $p_+(\gamma) = \Id$. Given $W_\gamma \in
\Gr^0_{(n)}$, we consider $\gamma$ as a map from $H_+^{(n)}$ to
$H^{(n)}$ by multiplication to the left. Then the point $W=W_\gamma$
can be described via a $\Z\times\N$ matrix representation of the map
$\gamma$ with respect to the bases of $H_+^{(n)}$ and of $H^{(n)}$
fixed in \eqref{basis}, that is,
\begin{equation}\label{Wwpm}
W\sim  %\left(\begin{array}{c}
%                    \om_-\\
%                    \om_+
%                    \end{array}\right) =
\Big(\gamma_{j-k}\Big)_{j \in \Z, k \in \N}:=
\left(\begin{array}{cccc}
\vdots &\vdots &\vdots & \iddots  \\
\gm_{-1}&\gm_{-2}&\gm_{-3}&\dots \\
\gm_0& \gm_{-1}&\gm_{-2}&\dots \\
\gm_1& \gm_0& \gm_{-1}&\dots \\
\vdots &\vdots &\vdots &  \ddots \\
    \end{array}\right),
\end{equation}
with blocks being the Fourier coefficients of $\gamma$.

Let $\om_\pm := p_\pm\circ\gamma$, these being maps
\[
\om_\pm: H_+^{(n)}\longrightarrow  H_\pm^{(n)}.
\]
We also introduce a map
\begin{equation}\label{}
h_W:H^{(n)}_+\to H^{(n)}_-
\end{equation}
whose graph is $W_\gamma$, namely,
\begin{equation}\label{}
h_W:=\om_-\circ \om_+^{-1}=p_-|_W\circ(p_+|_W)^{-1}
\end{equation}
(again we have used the property that $p_+|_W$ is one-to-one).

On the other hand, every element $g\in G_+^a$ defines a map by
multiplication
\[
g: H^{(n)} \longrightarrow H^{(n)}.
\]
Its inverse can be written in matrix form
%(the matrix
%representation becomes the transpose with respect to the second
%diagonal of that in \cite{SW},
(due to the basis \eqref{basis} we have fixed) as
%\footnote{The same remark as in the previous footnote applies.}
$$g^{-1}=\left(\begin{array}{ccccc}
                    d & 0\\
                    b & a
                    \end{array}\right)$$
where
\[
a:H^{(n)}_+\to H^{(n)}_+,\quad b: H^{(n)}_-\to H^{(n)}_+,\quad d:
H^{(n)}_-\to H^{(n)}_-.
\]

\begin{dfn}[\cite{SW}]\label{def-SWtau}
Given a point $W\in \Gr^0_{(n)}$, the associated tau function
depending on $g\in G_+^a$ is defined as
\begin{equation}\label{tauSSW}
\tau_{SSW}(g) := \det\Big(\Id+a^{-1} \circ b \circ h_W \Big) = \det\Big(\Id+b \circ h_W\circ a^{-1}\Big).
\end{equation}
\end{dfn}

\subsection{The related Riemann-Hilbert problem} \label{sec-HL}

Given a subspace $W_\gamma \in \Gr^0_{(n)}$,  we want to associate
to it a Riemann--Hilbert problem depending on $g \in G^{a}_+$.
%\textcolor{blue}{Chaozhong, because of the discussion about
%analyticity above we do not have to restrict to analytical loops
%here!}

Let us introduce the following matrix
\begin{equation}\label{jumpmatrix}
J_\gamma(g;z) := g^{-1}(z)\gamma(z), \quad z\in S^1.
\end{equation}
We choose an (infinite) set of coordinates $\bt $ on the abelian group
$G_+^a$, and consider the following Riemann-Hilbert problem with the
jump matrix $J_\gamma(\bt;z) = J_\gamma(g;z)$:
\begin{problem}\label{RH}
Find the (unique) sectionally-analytic function $\Gamma(\bt;z)$ on
$\C\setminus S^1$ such that
\begin{equation} \label{rh}
\left\{ \begin{array}{lll}
    \Gamma_+(\bt;z) &=& \Gamma_-(\bt;z)J_\gamma(\bt;z), \quad \forall z\in S^1,\\
    \\
    \Gamma_-(\bt;z) &\sim& \Id+\mathcal O(z^{-1}), \quad z \longrightarrow \infty.
    \end{array}\right.
\end{equation}
\end{problem}
Here the unit circle $S^1$ is oriented counter-clockwise,
$\Gamma_+(\bt;z)$ denotes the restriction of $\Gamma(\bt;z)$ to the
unit disk $D_0$ and $\Gamma_-(\bt;z)$ the restriction to its
complement.

\begin{prp}\label{thm-bfRH}
Given a Baker function $ w(\bt; z)$ for the point
$W_\gm=\gm(z)H_+^{(n)}\in\Gr^0_{(n)}$ acted by $G_+^a$, the function
\begin{equation}\label{Gam}
\Gamma_-(\bt; z) := w^{-1}(\bt; z)g(\bt;z)
\end{equation}
solves the Riemann-Hilbert problem~\ref{RH}. Conversely, suppose
that $\Gamma_-(\bt; z)$ is the solution of Problem~\ref{RH}, then
\begin{equation}\label{wRH}
 w(\bt; z) := g(\bt; z)\Gamma_-^{-1}(\bt; z)
\end{equation}
is a Baker function for the point $W_\gamma \in \Gr^0_{(n)}$. In
summary, given a point of the Grassmannian $\Gr^0_{(n)}$, the
Riemann--Hilbert problem \ref{RH} is solvable if and only if the
Baker function exists.
\end{prp}
\begin{prf}
Suppose $w(\bt;z)$ is a Baker function, then
$p_+(g^{-1}(\bt;z)w(\bt;z)) = \Id$, hence also
$$p_+[(g^{-1}(\bt;z)w(\bt;z))^{-1}] = \Id.$$ Moreover,
\begin{equation}\label{GmJ}
w^{-1}(\bt;z)g(\bt;z)\cdot g^{-1}(\bt;z)\gm(z)= w^{-1}(\bt;z)\gm(z)
\end{equation}
and the right hand side
$w^{-1}(\bt;z)\gm(z)=(\gm^{-1}(z)w(\bt;z))^{-1}$ belongs to
$H_+^{(n)}$; indeed we have $$w(\bt;z) \in \gamma H^{(n)}_+ =
W_\gamma$$ so that $\gamma^{-1}(z)w(\bt;z) \in H_+^{(n)}$. Observe
that the equality \eqref{GmJ} is just $\Gamma_-(\bt;z)
J_\gamma(\bt;z)=\Gamma_+(\bt;z)$. Thus the first assertion is
proven.

Conversely, given a solution of the Riemann--Hilbert problem, let us
check that the function $ w(\bt; z)$ in \eqref{wRH} satisfies the
two conditions in Definition~\ref{def-bf}. First, we have
\[
\gm^{-1}(z)w(\bt;z)=\gm^{-1}(z)g(\bt; z)\Gamma_-^{-1}(\bt;
z)=(\Gamma_-(\bt; z)J_\gamma(\bt;z))^{-1} =\Gamma_+^{-1}(\bt; z)\in
H_+^{(n)}.
\]
Hence $w(\bt;z)\in\gamma(z)H_+^{(n)}$. Second,
\begin{align}\label{}
%&g^{-1}(\bt;z)w(\bt;z) = \Gamma_-^{-1}(\bt; z),\\
&p_+(g^{-1}(\bt;z)w(\bt;z)) = p_+(\Gamma_-^{-1}(\bt; z) )=\Id.
\end{align}
Therefore the proposition is proven.
\end{prf}

Given a solution of the Riemann--Hilbert problem \eqref{rh}, the
so-called Malgrange form $\om_M$ is defined by (see \cite{Be} and
references therein)
\begin{equation}\label{omm}
\om_M(\p_t):=\oint_{S^1}\mathrm{Tr}\Big(\Gamma_-^{-1}(\bt;z)
\pd_z\Gamma_-(\bt;z)\cdot\Xi_t(\bt;z)\Big)\frac{d
z}{2\pi\mathbf{i}},
\end{equation}
where $\Xi_t(\bt;z)=\partial_{t}J_\gm(\bt;z)\cdot J_\gm(\bt;z)^{-1}$
for any $t\in\bt$.
\begin{lem}
The Malgrange form $\om_M$ for Riemann--Hilbert problem \eqref{rh}
is closed.
\end{lem}
\begin{prf}
The exterior differential of $\om_M$ is given in Proposition~2.1 in
\cite{Be}, which reads
\begin{equation}\label{omtt}
\p_{t'}\om(\p_t)-\p_{t}\om(\p_{t'})=\frac{1}{2}\oint_{S^1}
\mathrm{Tr}\Big(\Xi_{t'}(\bt;z)\p_z\Xi_t(\bt;z)-\p_z\Xi_{t'}(\bt;z)\cdot\Xi_t(\bt;z)\Big)\frac{d
z}{2\pi\mathbf{i}}
\end{equation}
for any $t,t'\in\bt$. In the present case, we have
\[
\Xi_t(\bt;z)=\p_t(g^{-1}(\bt;z)\gm(z))\cdot\gm^{-1}(z)g(\bt;z)
=-g^{-1}(\bt;z)\p_t g(\bt;z),
\]
which is analytic on the whole complex plane. Thus the right hand
side of \eqref{omtt} vanishes. The lemma is proved.
\end{prf}

\begin{dfn}[\cite{Be}]
The (generalized) Jimbo-Miwa-Ueno isomonodromic tau function for the
Riemann--Hilbert problem \ref{RH}, supposed solvable, is defined up
to a constant factor
%as the unique functions whose derivatives are given
by
\begin{equation}\label{tauJMU}
\partial_{t}\log\tau_{JMU}(\bt)=\oint_{S^1}\mathrm{Tr}\Big(\Gamma_-^{-1}(\bt;z)
\pd_z\Gamma_-(\bt;z)\cdot\partial_{t}J_\gm(\bt;z)\cdot
J_\gm^{-1}(\bt;z)\Big)\frac{d z}{2\pi\mathbf{i}}
\end{equation}
for any $t\in\bt$.
\end{dfn}

\begin{rmk}
In \cite{Be} a tau function was introduced by Bertola for a general
Riemann--Hilbert problem on the Riemann sphere, under very mild assumptions. He also
showed that, whenever the Riemann--Hilbert problem is associated to
the isomonodromy data of an ODE, the tau function coincides with the
isomonodromy tau function of Jimbo, Miwa and Ueno \cite{JM1, JM2,
JMU}. This is why we use the notation $\tau_{JMU}$.
\end{rmk}

In consideration of the equivalence between the Baker function and
the solution of the Riemann--Hilbert problem, it is natural to study
the relation between the Sato--Segal--Wilson and the
Jimbo--Miwa--Ueno isomonodromy tau function. This is what we plan to
do in the next section, identifying both tau functions with the
Szeg\"o--Widom pre-factor $D_{\infty}(J_\gamma)$ (see below). As a
result, we will obtain the following theorem:

\begin{thm}\label{main}
Given a point $W=\gamma H_+^{(n)}\in\Gr^0_{(n)}$ acted by a group
$G_+^a$, the Sato--Segal--Wilson tau function $\tau_{SSW}(\bt)$
defined in \eqref{tauSSW} coincides (up to constants) with the
isomonodromic tau function $\tau_{JMU}(\bt)$ associated to the
Riemann--Hilbert problem \ref{RH}.
\end{thm}

In combination of Proposition~\ref{thm-bfRH} and Theorem~\ref{main},
we obtain immediately
\begin{cor}\label{thm-tauw}
Given a point $W=\gamma H_+^{(n)}\in\Gr^0_{(n)}$ acted by
$g(\bt;z)\in G_+^a$, the Baker function and the Sato--Segal--Wilson
tau function are related by
\begin{equation}\label{bftau}
\partial_{t}\log\tau_{SSW}(\bt)=\oint_{S^1}\mathrm{Tr}\Big(
\pd_z w(\bt;z)\cdot w^{-1}(\bt;z)\pd_t g(\bt;z)\cdot g^{-1}(\bt;z)
\Big)\frac{d z}{2\pi\mathbf{i}}
\end{equation}
for any $t \in \bt$ of $G_+^a$.
\end{cor}
\begin{prf}
One substitutes \eqref{Gam} and \eqref{jumpmatrix} into
\eqref{tauJMU}, then gets
\begin{align}\label{bftau2}
\partial_{t}\log\tau_{SSW}(\bt)=\oint_{S^1}\mathrm{Tr}\Big(
\pd_z( g^{-1}(\bt;z)w(\bt;z) )\cdot w^{-1}(\bt;z)g(\bt;z)\cdot
g^{-1}(\bt;z)\pd_t g(\bt;z) \Big)\frac{d z}{2\pi\mathbf{i}}.
\end{align}
The equality is converted to \eqref{bftau} by using that $g(\bt;z)$
is analytic for all $z\in\C$. The lemma is proved.
\end{prf}

\begin{rmk}
The equalities \eqref{bftau2} or \eqref{bftau} can be understood as
some generalization of the Sato formula that connects Baker function
and tau function in scalar case. For instance, the Sato formula for
the KP hierarchy reads
\begin{equation}\label{wKP}
w(\bt;z)=g(\bt;z)\frac{\tau(\bt-[z^{-1}])}{\tau(\bt)}, \quad
g(\bt;z)=\exp\left(\sum_{k=1}^\infty t_k z^k\right),
\end{equation}
where $\bt=(t_1, t_2, t_3,\dots)$ and $[z]=(z/1, z^2/2,
z^3/3,\dots)$. It is easy to check that this formula agrees with the
equality \eqref{bftau2}.
%One ``substitutes'' \eqref{wKP} into the right hand
%side of \eqref{bftau2} and have
%\begin{align}\label{}
%\mathrm{r.h.s.}=&\oint_{S^1} \pd_z
%\frac{\tau(\bt-[z^{-1}])}{\tau(\bt)}\cdot
%\frac{\tau(\bt)}{\tau(\bt-[z^{-1}])} g^{-1}(\bt;z) \pd_{t_k}
%g(\bt;z)
%\frac{d z}{2\pi\mathbf{i}} \nn\\
%=& \oint_{S^1}
%\pd_z\log\tau(\bt-[z^{-1}])\cdot z^k\frac{d z}{2\pi\mathbf{i}} \nn\\
%=& \oint_{S^1} \sum_{l=1}^\infty\pd_{t_l}\log\tau(\bt)z^{-l-1}\cdot
%z^k\frac{d z}{2\pi\mathbf{i}} \nn\\
%=& \pd_{t_k}\log\tau(\bt),
%\end{align}
%which agrees with \eqref{bftau2}, or equivalently, \eqref{bftau}.
%From this example, we also see that usually it is more convenient to
%apply the formula \eqref{bftau2} rather than  \eqref{bftau}.
\end{rmk}

An immediate application of Corollary~\ref{thm-tauw} is to deduce a
connection between the Sato--Segal--Wilson tau function and the
matrix Baker function of the AKNS-D hierarchy in \cite{Dickey}. What
is more, the formula \eqref{bftau2} will be used to study the
relationship of $\tau_{SSW}$ for Drinfeld--Sokolov hierarchies and
their tau functions introduced in \cite{Wu}, see
Section~\ref{sec-DS} below.

\section{Tau functions and Toeplitz determinants} \label{sec-td}

In this section let us proceed to prove Theorem~\ref{main}.

\subsection{Toeplitz determinant and the Szeg\"o-Widom theorem}
First we digress to review some results on large-size block Toeplitz
determinants to be used in the present article. These results, due
to Widom, can be found in \cite{Wi74, Wi75, Wi76}. Some of our
notations are borrowed from \cite{BotSilb}.

Given a loop $\varphi=\sum_{j\in\Z}\varphi_j z^j \in L_{1/2}GL_n$,
it is easy to see that the operator of multiplication
\[
\varphi: H^{(n)} \longrightarrow H^{(n)},
\]
for the basis \eqref{basis}, has (block) matrix representation given
by the \emph{Laurent} matrix $L(\varphi) :=
(\varphi_{j-k})_{j,k\in\Z}$. In the sequel, given $\varphi \in
L_{1/2}U_n$, we are interested in the following associated $\N
\times \N$ matrices:
\begin{equation}
    T(\varphi) := \Big(\varphi_{j-k}\Big)_{j,k \in \N}; \quad H(\varphi) := \Big(\varphi_{j+k+1}\Big)_{j,k \in \N}; \quad \widetilde H(\varphi) := \Big(\varphi_{-j-k-1}\Big)_{j,k \in \N}.
\end{equation}
The first matrix is the so-called (block) Toeplitz matrix associated
to the symbol $\varphi$, while the second and the third are the two
Ha\"enkel matrices associated to $\varphi$.

\newpage

Let us introduce the following involution operator
\begin{align}\label{}
\iota: H^{(n)} &\longrightarrow H^{(n)} \nn\\
v(z) & \longmapsto v(z^{-1})z^{-1}.
\end{align}
Clearly, $\iota\circ \iota=\Id$, and the restrictions $\iota:
H^{(n)}_\pm \rightarrow H^{(n)}_\mp$ are one-to-one correspondences.
 It is easy to check that the matrices above are the matrix
representations of the following operators (endomorphisms of
$H^{(n)}_+$):
\begin{equation}\label{oprepresentation}
    T(\varphi) = p_+ \circ \varphi{|_{H_+^{(n)}} }; \quad H(\varphi) = p_+ \circ \varphi \circ \iota{|_{H_+^{(n)}} }; \quad \widetilde H(\varphi) = \iota\circ p_- \circ \varphi{|_{H_+^{(n)}} }.
\end{equation}
Using \eqref{oprepresentation} it turns out to be a convenient way
to write identities between $\N \times \N$ matrices, as the
following useful lemma will show (see, for instance,
\cite{BotSilb}):
\begin{lem}
    Given $\varphi_1,\varphi_2 \in L_{1/2}GL_n$, we have the following identity between $\N \times \N$ matrices:
    \begin{equation}\label{toeplitzhaenkelidentity}
        T(\varphi_1)T(\varphi_2) = T(\varphi_1\varphi_2) - H(\varphi_1)\widetilde H(\varphi_2).
    \end{equation}
\end{lem}
\begin{prf}
 Starting from the left hand side, we have
 \begin{eqnarray}\nonumber
    T(\varphi_1)T(\varphi_2) &=& p_+ \circ \varphi_1 \circ p_+ \circ \varphi_2|_{H_+^{(n)} }
    = p_+ \circ \varphi_1  \circ (\Id - p_-) \circ \varphi_2|_{H_+^{(n)}}
    \\ &=& p_+ \circ (\varphi_1\varphi_2){|_{H_+^{(n)}} } -
    p_+\circ\varphi\circ \iota\circ \iota \circ p_- \circ \varphi_2|_{H_+^{(n)} } \nn\\
    &=& T(\varphi_1\varphi_2) - H(\varphi_1)\widetilde H(\varphi_2).
 \end{eqnarray}
 Thus the lemma is proved.
\end{prf}

Also, from the definition of $L_{1/2}GL_n$ it follows that both
$H(\varphi)$ and $\widetilde H(\varphi^{-1})$ are Hilbert--Schmidt
operators; hence their product is a trace-class operator and,
consequently, the Fredholm determinant of the operator $\Id -
H(\varphi)\widetilde H(\varphi^{-1})$ is well defined (see for
instance \cite{Si}).

Now let  $T_N(\varphi)$ denote the $(N+1)\times(N+1)$ upper-left
principal minor of $T(\varphi)$, that is,
\[
T_N(\varphi):= \left(\begin{array}{cccc}
\varphi_0 & \varphi_{-1} & \ldots & \varphi_{-N}\\
&&&\\
\varphi_1 & \varphi_0 & \ldots & \varphi_{-N+1}\\
&&&\\
   \vdots       & \vdots & \ddots & \vdots\\
&&&\\
\varphi_N & \varphi_{N-1} & \ldots & \varphi_0
\end{array}\right),
\]
whose determinant is denoted as $D_N(\varphi)$. Below is the
celebrated Szeg\"o--Widom theorem.
\begin{thm}[Szeg\"o--Widom theorem \cite{Wi76}]
\label{thm-SW} Assume $\varphi \in L_{1/2}GL_n$. Then it exists the
large-size limit
\begin{equation}\label{}
D_\infty(\varphi):=\lim_{N\to\infty}
\frac{D_N(\varphi)}{G(\varphi)^{N+1}}=\det\Big(T(\varphi)T(\varphi^{-1})\Big) = \det\Big(\Id - H(\varphi)\widetilde H(\varphi^{-1})\Big),
\end{equation}
where
\[
G(\varphi)=\exp\left(\frac{1}{2\pi}\int_0^{2\pi}\log\big(\det
\varphi(e^{\mathbf{i}\theta})\big)d\theta\right).
\]
\end{thm}

Let us also recall a result, again due to Widom (and rederived in \cite{IJK} with a different approach), connecting the
Riemann--Hilbert factorizations of the symbol $\varphi$ with
$D_\infty(\varphi)$.
\begin{thm}[Theorem~4.1 in \cite{Wi74}]\label{factorization}
 Suppose that a loop $\varphi$ satisfies the conditions imposed in the Szeg\"o--Widom
 theorem, and moreover it depends in a differentiable way on a given parameter $t$.
 If $\varphi^{-1}(z)$ admits two Riemann--Hilbert factorizations
\begin{equation}\label{}
\varphi^{-1}(z)=T_+(z)T_-(z)=S_-(z)S_+(z), \quad  z\in S^1
\end{equation}
with
 \bea
    T_+(z):=\sum_{k\geq 0}(T_+)_{k}z^k,\quad\quad S_+(z):=\sum_{k\geq 0}(S_+)_{k}z^k,\nonumber\\
    T_-(z):=\sum_{k\leq 0}(T_-)_{k}z^k,\quad\quad S_-(z):=\sum_{k\leq 0}(S_-)_{k}z^k,\nonumber
 \eea
then
 \begin{equation}\label{RH Widom}
        \pd_t\log(D_\infty(\varphi))=
        -\oint_{S^1} \mathrm{Tr}\left[\Big((\partial _z T_+)T_--(\partial _z S_-)S_+\Big)
        \partial _t \varphi\right]\frac{d z}{2\pi\mathbf{i}}.
    \end{equation}
\end{thm}

\subsection{Tau functions as the large-size limit of block Toeplitz
determinants}

Let us come back to the setting in Section~\ref{sec-SSW}. For any
point $W_\gamma = \gamma(z)H^{(n)}_+\in\Gr_0^{(n)}$ acted by
$g(\bt;z)\in G_+^{a}$, one has the following jump matrix for the
related Riemann--Hilbert problem
\begin{equation}\label{}
J_\gamma(\bt;z):=g^{-1}(\bt;z)\gamma(z), \quad z\in S^1.
\end{equation}
The first step to prove Theorem~\ref{main} is the following
\begin{thm}\label{thm-DM}
The large-size limit of the block Toeplitz determinant
$D_\infty(J_\gamma(\bt;z))$ coincides with the corresponding
Sato--Segal--Wilson tau function defined by \eqref{tauSSW}, that is,
\begin{equation}\label{}
D_\infty(J_\gamma(\bt;z))=\tau_{SSW}(\bt).
\end{equation}
\end{thm}
\begin{prf}
By the very definition in Section~\ref{sec-SSW}, we have
\[
\tau_{SSW}(\bt) = \det(\Id + b\circ\omega_-\circ\omega_+^{-1}\circ
a^{-1})
\]
where the terms in the determinant can be written as follows:
\begin{eqnarray}
    &a^{-1} = p_+ \circ g|_{H^{(n)}_+};& b = p_+ \circ g^{-1}|_{H^{(n)}_-} ;\\
    &\omega_+^{-1} = p_+ \circ \gamma^{-1}|_{H^{(n)}_+};& \omega_- = p_- \circ \gamma{|_{H_+^{(n)}} }.
\end{eqnarray}
Hence we have \footnote{Here and below we suppress the sign of
composition.}
\begin{equation}\label{eqn1}
    \tau_{SSW} = \det\Big(\Id + p_+  g^{-1}  p_-  \gm \;  p_+  \gm^{-1} g|_{H_+^{(n)} }\Big).
\end{equation}
On the other hand, by virtue of Theorem~\ref{thm-SW} and the
relations \eqref{oprepresentation} we have
\begin{eqnarray}
    D_\infty(J_\gamma) &=& \det(T(J_{\gamma})T(J_{\gamma}^{-1})) \nn\\
    & =& \det \Big(p_+  g^{-1}  \gamma \; p_+  \gamma^{-1}  g{|_{H_+^{(n)}} } \Big)\nonumber \\
    &=& \det\Big(\Id - p_+  g^{-1}  \gamma \; p_-  \gamma^{-1} g{|_{H_+^{(n)}} }\Big). \label{eqn2}
\end{eqnarray}
Combining \eqref{eqn1} and \eqref{eqn2}, it is sufficient to prove
\begin{equation}\label{eqn3}
    \left.\left(p_+  g^{-1}  p_-  \gm \; p_+  \gm^{-1}  g + p_+  g^{-1}
    \gamma \; p_-  \gamma^{-1} g\right)\right|_{H_+^{(n)}}  = 0.
\end{equation}
Indeed, the left hand side is
\begin{align}\label{}
\mathrm{l.h.s.}=& \left.\left(-p_+  g^{-1}  p_+  \gm \; p_+ \gm^{-1}
g + p_+  g^{-1}  \gm \; p_+ \gm^{-1} g + p_+  g^{-1}
    \gamma \; p_-  \gamma^{-1} g\right)\right|_{H_+^{(n)}} \nn\\
=& \left.\left(-p_+  g^{-1} \om_+  \; \om_+^{-1} g + p_+ g^{-1} \gm
\;  \gm^{-1} g \right)\right|_{H_+^{(n)}} \nn\\
=&\left.\left(-\Id + \Id\right)\right|_{H_+^{(n)}}=0.
\end{align}
%Substituting, in both addendum, $p_-$ with $\Id - p_+$ we get that
%\eqref{eqn3} is true if and only if
%$$\Id = p_+ g^{-1} p_+ \gamma p_+ \gamma^{-1} g{|_{H_+^{(n)}} } = p_+ g^{-1} p_+ \gamma p_+ \gamma^{-1} p_+ g{|_{H_+^{(n)}} };$$
%and using the fact that, because of  \eqref{toeplitzhaenkelidentity},
%\begin{eqnarray}
%    p_+ \gamma \; p_+ \gamma^{-1}{|_{H_+^{(n)}} } = \Id; \quad p_+ g^{-1} p_+ g{|_{H_+^{(n)}} } = \Id \nonumber,
%\end{eqnarray}
Thus we conclude the theorem.
\end{prf}

The equivalence between the Sato--Segal--Wilson tau functions and
certain large-size block Toeplitz determinants was established, for
the case of Gelfand--Dickey hierarchies, by one of the authors in
\cite{Ca}. The proof in the present article is completely different
in nature. Indeed, it is more general and much more straightforward.
On the other hand, as a byproduct of the proof in \cite{Ca}, we
obtained that also each $D_N(J_\gamma)$, and not just the limit for
$N \rightarrow \infty$, is a tau function. In this article, no
similar statement is made about $D_N(J_\gamma)$.

Let us now study the relationship between $D_\infty(J_\gamma)$ and
the Jimbo--Miwa--Ueno isomonodromy tau function.
\begin{thm}\label{thm-JMU}
Up to a constant factor, the large-size limit of the block Toeplitz
determinant $D_\infty(J_\gm(\bt;z))$ coincides with the
Jimbo--Miwa--Ueno isomonodromy tau function defined by
\eqref{tauJMU}, that is,
\begin{equation}\label{}
D_\infty(J_\gm(\bt;z))=\tau_{JMU}(\bt).
\end{equation}
\end{thm}
\begin{prf}
We need to show
\begin{equation}\label{DtauJMU}
 \partial_t\log D_\infty(J_\gm(\bt;z))=\partial_t\log\tau_{JMU}(\bt)
\end{equation}
for any $t$ in the coordinates $\bt$ of the group $G_+^a$.

The inverse of $J_\gm$ has the following two Riemann--Hilbert
factorizations
\begin{align}\label{}
&J^{-1}_\gamma=\gamma^{-1}(z)g(\bt;z)=S_-S_+,\\
&J^{-1}_\gamma=\Gamma_+^{-1}(\bt;z)\Gamma_-(\bt;z)=T_+T_-.
\end{align}
According to Theorem~\ref{factorization}, we have (here the prime
denotes the derivative with respect to $z$, and $J=J_\gm$ to avoid
lengthy notations)
\begin{align}\label{DJt}
 \partial_t\log D_{\infty}(J)&=\oint\mathrm{Tr}
 \Big((\Gamma_+^{-1}\Gamma_+'\Gamma_+^{-1}\Gamma_-
 -\gamma^{-1}\gamma'\gamma^{-1}g)\partial_t
 J\Big)\frac{dz}{2\pi\mathbf{i}}.
\end{align}
Let $I_1-I_2$ denote the right hand side of \eqref{DJt}. We have
\begin{align}
 I_1=&\oint\mathrm{Tr}
 \Big(\Gamma_+^{-1}(\Gamma_-'J+\Gamma_- J')\Gamma_+^{-1}\Gamma_-\partial_t
 J\Big)\frac{dz}{2\pi\mathbf{i}} \nn\\
=&\oint\mathrm{Tr}
 \Big(J^{-1}\Gamma_-^{-1}\Gamma_-'\partial_t
 J\Big)\frac{dz}{2\pi\mathbf{i}} + \oint\mathrm{Tr}\Big(J^{-1}J' J^{-1}\partial_t
 J\Big)\frac{dz}{2\pi\mathbf{i}} \label{3.19},
\end{align}
in which the second term reads
\begin{align}\label{}
 -\oint\mathrm{Tr}\Big((J^{-1})'\partial_t
 J\Big)\frac{dz}{2\pi\mathbf{i}} =&
 -\oint\mathrm{Tr}\Big((-\gm^{-1}\gm'\gm^{-1}g+\gm^{-1}g')\partial_t
 J\Big)\frac{dz}{2\pi\mathbf{i}} \nn\\
 =& I_2-\oint\mathrm{Tr}\Big(\gm^{-1}g'\partial_t g^{-1}\cdot\gm\Big)
 \frac{dz}{2\pi\mathbf{i}} \nn\\
 =&I_2-0.
\end{align}
Thus
\begin{equation}\label{}
 \partial_t\log D_{\infty}(J)=I_1-I_2=\oint\mathrm{Tr}
 \Big(J^{-1}\Gamma_-^{-1}\Gamma_-'\partial_t
 J\Big)\frac{dz}{2\pi\mathbf{i}},
\end{equation}
which coincides with $\pd_t\log\tau_{JMU}$ by the definition in
\eqref{tauJMU}, hence the equality \eqref{DtauJMU} is valid.
Therefore the theorem is proved.
\end{prf}

Finally, taking Theorems~\ref{thm-DM} and \ref{thm-JMU} together, we
complete the proof of Theorem~\ref{main}.

\section{Tau functions of generalized Drinfeld--Sokolov hierarchies}
\label{sec-DS}

In \cite{DS} Drinfeld and Sokolov constructed a Hamiltonian
integrable hierarchy of  Korteweg--de Vries (KdV) type associated to
every affine Kac--Moody algebra. Provided a solution of the
Drinfeld--Sokolov hierarchy, we want to give a Grassmannian
construction for its tau function, and compare it with other tau
functions introduced in the literature. In the same way, tau
functions of generalized Drinfeld--Sokolov hierarchies proposed by
Groot, Hollowood and Miramontes \cite{GHM, HM} will also be
considered.

\subsection{Affine Kac--Moody algebras and Drinfeld--Sokolov hierarchies}

%Firstly, we lay out some preliminary results on Kac--Moody algebras.

Let $A=(a_{i j})_{0\le i, j\le l}$ be a Cartan matrix of affine type
$X_N^{(r)}$, and $\fg(A)$ be the corresponding Kac--Moody algebra.
This algebra can be realized as follows.

Introduce a set of integer vectors (known as gradations):
\begin{equation}\label{}
\Gm=\{(s_0, s_1, \dots, s_l)\in\Z^{l+1}\mid s_i\ge0,
s_0+s_1+\cdots+s_l>0\}.
\end{equation}
For example, the  following vectors
\begin{equation}\label{}
\rs^0=(1,0,\dots,0), \quad \rs^1=(1,1,\dots,1)
\end{equation}
are called the homogeneous and the principal gradations
respectively. Given an arbitrary vector $\rs=(s_0, s_1, \dots,
s_l)\in\Gm$, denote $N_{\rs}=\sum_{i=0}^l k_i s_i$, where $k_i$ are
the Kac labels of $\fg(A)$, i.e., the lowest positive integers
solving $\sum_{j=0}^la_{i j}k_j=0$. Note, in particular, that
$k_0=1$ in all cases except $k_0=2$ for the case of type
$A_{2l}^{(2)}$.

Let $\mathcal{G}$ be the simple Lie algebra of type $X_N$, whose
Dynkin diagram has an automorphism of order $r$. The Lie algebra
$\mathcal{G}$ is generated by certain special elements $E_i$, $F_i$
and $H_i$ ($i=0, 1, \dots, l$) as presented in \S\,8.3 of
\cite{Kac}. Given an integer vector $\rs\in\Gm$, it induces a $\Z/r
N_\rs\Z$-gradation
\[
 \mathcal{G}=\bigoplus_{k=0}^{r
N_\rs-1}\mathcal{G}_k
\]
 by assigning
\[
\deg E_i=s_i, \quad \deg F_i=-s_i,\quad  \deg H_i=0, \qquad 0\le
i\le l.
\]
With the help of a parameter $z$, the affine Kac--Moody algebra
$\fg(A)$ graded by $\rs$ can be realized as (in the present paper
the scaling element $d$ is not needed)
\begin{equation}\label{gAs}
\fg(A;\rs)=\bigoplus_{k\in\Z}\left(
z^k\otimes\mathcal{G}_{k\!\!\mod\,r N_{\rs}}\right)\oplus \C\,c.
\end{equation}
Here $c$ is the canonical central element, and the Lie bracket
between elements of the form $X(k), Y(k)\in
z^k\otimes\mathcal{G}_{k\!\!\mod\,r N_{\rs}}$ is defined by
\begin{align}\label{XYbr}
&[X(j), Y(k)]=[X, Y](j+k)+\dt_{j,-k}\frac{j}{r N_\rs}(X\mid Y)_0\,c,
\end{align}
with $(\,\cdot\mid\cdot\,)_0$ being the standard invariant symmetric
bilinear form on $\mathcal{G}$.

A set of Weyl generators of $\fg(A;\rs)$ can be chosen as
\begin{equation}\label{weyls}
e_i^{(\rs)}=E_i(s_i), \quad  f_i^{(\rs)}=F_i(-s_i), \quad
\al_i^{\vee(\rs)}=H_i(0)+\frac{k_i s_i}{k^\vee_i N_\rs}c,
\end{equation}
where $k_i$ and $k^\vee_i$ are the Kac labels and the dual Kac
labels respectively, and $i=0,1,\dots,l$. Under the gradation $\rs$,
we have
\begin{equation}\label{dgr}
\deg e_i^{(\rs)}=s_i, \quad  \deg f_i^{(\rs)}=-s_i, \quad
\deg\al_i^{\vee(\rs)}=0, \qquad 0\le i\le l.
\end{equation}
In particular, the elements $\al_i^{\vee(\rs)}$ satisfy
$\sum_{i=0}^l k_i^\vee\al_i^{\vee(\rs)}=c$ and they generate the
Cartan subalgebra of $\fg(A;\rs)$.

Given any other gradation $\rs'\in\Gm$, there is a natural
isomorphism between $\fg(A;\rs)$ and $\fg(A;\rs')$ induced by
\[
e_i^{(\rs)}\mapsto e_i^{(\rs')}, \quad  f_i^{(\rs)}\mapsto
f_i^{(\rs')}.
\]
Henceforth we simply write $\fg=\fg(A)$ instead of $\fg(A;\rs)$. We
also write $\fg=\bigoplus_{j\in\Z}\fg_{j\,[\rs]}$ graded by $\rs$,
and use subscript ``$<0\,[\rs]$'' to stand for the projection
$\mathrm{pr}: \fg\to \bigoplus_{j<0}\fg_{j\,[\rs]}$, e.t.c.

A key role in the construction of Drinfeld--Sokolov hierarchies
\cite{DS} is played by the principal Heisenberg subalgebra
$\mathcal{H}$ of $\fg$. More exactly, let $E$ be the set of
exponents of $\fg$, then there are elements
$\Ld_j\in\fg_{j\,[\rs^1]}$ for $j\in E$ such that
\[
\cH_{[\rs']}=\C c\oplus\sum_{j\in E}\C\Ld_j,
\]
with
\begin{equation}\label{}
[\Ld_j, \Ld_k]=j\dt_{j,-k}c, \quad j, k\in E.
\end{equation}
In particular, $1$ is an exponent for every affine Kac--Moody
algebra (see \cite{Kac}), and $\Ld=\Ld_1$ is a semisimple element.
Namely,
\begin{equation}\label{dec2}
\fg=\cH+\im\,\ad_{\Ld}, \quad \cH\cap\im\,\ad_{\Ld}=\C c;
\end{equation}
note $\mathrm{ker}\,\ad_\Ld=\cH$ modulo the center.

Introduce an operator
\begin{equation}\label{sLgp}
\sL=\frac{\od}{\od x}+\Ld+q,
\end{equation}
where $q$ is a smooth function of $x\in\R$ taking value in
$\left.\left(\fg_{0\,[\rs^0]}\cap\fg_{\le0\,[\rs^1]}\right)\right/\C
c$. By using the property \eqref{dec2}, one has the following
dressing proposition.
\begin{prp}[\cite{DS}]\label{thm-dr0}
There exists a smooth function $U\in C^\infty(\R,
\fg_{<0\,[\rs^1]})$ such that the operator $\bar{\sL}=e^{-\ad_U}\sL$
has the form
\begin{equation}\label{}
\bar{\sL}=\frac{\od}{\od x}+\Ld+H, \quad H\in C^\infty(\R,
\mathcal{H}_{<0\,[\rs^1]}\oplus\C\,c).
\end{equation}
%Moreover, both $U$ and $H$ are differential polynomials in (components of) $q$.
\end{prp}

On $\fg$ the Drinfeld--Sokolov hierarchy is defined by the following
partial differential equations
\begin{equation}\label{Lt2}
\frac{\pd \sL}{\pd t_j}=[(e^{\ad_U}\Ld_j)_{<0\,[\rs^0]}, \sL], \quad
j\in E_{>0}
\end{equation}
restricted to some equivalence class of $\sL$ with respect to the
gauge actions
\begin{equation}\label{gauge}
\sL\mapsto \tilde{\sL}=e^{\ad_T}\sL,  \quad T\in C^\infty(\R,
\fg_{0\,[\rs^0]}\cap\fg_{<0\,[\rs^1]}).
\end{equation}
One can choose a special gauge slice of the operator $\sL$ such that
(see \cite{HM, Mi}) it satisfies
\begin{equation}\label{LTa}
\sL=e^{\ad_V}\left(\frac{\od}{\od x}+\Ld\right)+f\cdot c,
%=\Ta\left(\frac{\od}{\od x}+\Ld\right)\Ta^{-1}
\end{equation}
where $V$ is a function taking value in $\fg_{<0\,[\rs^0]}$ and $f$
is a scalar function. Let $\Ta=e^V$, i.e., an element in the
Kac--Moody group of $\fg$, then the Drinfeld--Sokolov hierarchy
\eqref{Lt2} can be written equivalently as
\begin{equation}\label{Tat}
\frac{\pd\Ta}{\pd t_j}=(\Ta\Ld_j\Ta^{-1})_{<0\,[\rs^0]}\,\Ta, \quad
j\in E_{>0}.
\end{equation}

We remark that both functions $U$ and $H$ in
Proposition~\ref{thm-dr0} are differential polynomials in $q$, but
$V$ above may not be a differential polynomial in $q$. An algorithm
to calculate $U$, $H$ and $V$ was given by one of the authors in
\cite{Wu}, and these functions define a tau function that is
independent of the choice of gauge equivalent class of $\sL$.
\begin{dfn}[\cite{Wu}]\label{}
The tau function $\tau(\bt)$ of the Drinfeld--Sokolov hierarchy
\eqref{Lt2} is defined by
\begin{equation}\label{tauDS}
\pd_{t_j}\log\tau=-(\Ta \Ld_j\Ta^{-1})_c, \quad j\in E_{>0},
\end{equation}
where the subscript ``$c$'' means the coefficient of $c$ with
respect to the following decomposition of the Cartan subalgebra of
$\fg$:
\[
\mathfrak{h}=\C\al_1^\vee\oplus\dots\oplus\C\al_l^\vee\oplus\C\,c.
\]
\end{dfn}

\subsection{Tau functions of
Drinfeld--Sokolov hierarchies}

Henceforth we identify $\mathcal{G}$ with its realization by
$n\times n$ trace-less matrices as in \cite{Kac, DS}. In this case,
the standard invariant bilinear form reads
\[
(X\mid Y)_0=\kappa\, \mathrm{Tr}(X Y), \quad X, Y\in\mathcal{G}
\]
with some constant $\kappa$. For instance, $\kappa=1$ for the
special linear/sympletic algebras (types A and C) and $\kappa=1/2$
for the special orthogonal algebras (types B and D), see the
appendix of \cite{DS}. Accordingly, we realize $\fg=\fg(A;\rs^0)$ as
\eqref{gAs} with $\rs^0$ being the homogeneous gradation.

Since $\mathcal{G}$ is realized by trace-less matrices, we consider
$z$ in \eqref{gAs} as a complex parameter, and take the following
two subgroups of $L_{1/2}U_n$:
\begin{align}\label{lieGm}
&G_-=\left\{ e^X\in L_{1/2}U_n\mid X\in\fg_{<0\,[\rs^0]}\right\}, \\
&G_+^a=\left\{g(\bt ; z)=\exp\left(\sum_{j\in E_{>0}}t_j
\Ld_j\right)\in L_{1/2}U_n\mid t_j\in\R \right\}. \label{lieGa}
\end{align}
Observe that under the realization of $\fg$, every element of $G_-$
takes the form $\Id+\mathcal{O}(z^{-1})$, and that $G_+^a$ is an
abelian group of functions holomorphic on the complex plane.

Assume  $\Ta(\bt;z)$ to be an arbitrary solution of the
Drinfeld--Sokolov hierarchy \eqref{Tat}. Clearly,
$\Ta(\bt;z)=I+O(z^{-1})$ is a function taking value in the Lie group
$G_-$. We introduce
\begin{equation}\label{bakerTa}
 w(\bt; z)=g(\bt; z)\Ta^{-1}(\bt; z), \quad g(\bt; z)\in G_+^a.
\end{equation}

\begin{prp}\label{thm-bakerDS}
Given a point $W= \gm(z)H_+^{(n)}\in\Gr^0_{(n)}$ with
$\gm(z)=\Ta^{-1}(0;z)\in G_-$, the function $w(\bt; z)$ is the
corresponding Baker function that depends on the parameter $\bt$ of
$G_+^a$ and $z\in S^1$.
\end{prp}
\begin{prf}
To simply notations, we will use subscripts ``$\pm$'' to stand for
the projections $p_\pm$ given in Section~\ref{sec-SSW}, which is
consistent with the following decomposition of affine Kac--Moody
algebra
\[
\fg=\fg_{\ge0\,[\rs^0]}\oplus\fg_{<0\,[\rs^0]}.
\]

Recalling Definition~\ref{def-bf}, clearly $w(\bt; z)$ satisfies the
second condition, namely
\[
(g^{-1}(\bt; z)w(\bt; z))_+=\Ta^{-1}(\bt; z)_+=\Id.
\]
For the first condition, we only need to show that
\begin{equation}\label{}
\hat{w}(\bt; z):=\gm^{-1}(z)w(\bt; z)
\end{equation}
belongs to $H_+^{(n)}$. In fact, for every $j\in E_{>0}$, one has
\begin{align}
\pd_{t_j} w =& g \Ld_j
\Ta^{-1} - g\Ta^{-1} \pd_{t_j}\Ta\cdot\Ta^{-1}   \nn\\
=&g\Ta^{-1} \cdot\Ta \Ld_j \Ta^{-1} - g\Ta^{-1} (\Ta\Ld_j\Ta^{-1})_- \nn\\
=&w (\Ta\Ld_j\Ta^{-1})_+, \nn\\
\end{align}
hence
\begin{equation}\label{}
\hat{w}^{-1}\pd_{t_j}\hat{w}=w^{-1}\pd_{t_j} w
=(\Ta\Ld_j\Ta^{-1})_+\in\fg_{\ge0\,[\rs^0]}.
\end{equation}
This together with the initial value
$\left.\hat{w}\right|_{\bt=0}=\Id$ implies that the function
$\hat{w}$ takes value in the Lie group of $\fg_{\ge0\,[\rs^0]}$;
namely, $\hat{w}$ contains only nonnegative powers in $z$. Therefore
the proposition is proved.
\end{prf}

Given the point $W=\gm(z)H_+^{(n)}$ in the Grassmannian, it is
defined the Sato--Segal--Wilson tau function $\tau_{SSW}(\bt)$
(recall Definition~\ref{def-SWtau}). Equivalently, this tau function
is given by the generalized Sato formula \eqref{bftau2} as
\begin{equation}\label{satoDS}
\partial_{t_j}\log\tau_{SSW}(\bt)=\oint_{S^1}\mathrm{Tr}\Big(\pd_z\Ta^{-1}\cdot
\Ta \Ld_j \Big)\frac{d
z}{2\pi\mathbf{i}}=-\oint_{S^1}\mathrm{Tr}\Big(\pd_z\Ta\cdot \Ld_j
\Ta^{-1} \Big)\frac{d z}{2\pi\mathbf{i}}
\end{equation}
with $j\in E_{>0}$. In other words, we obtain a tau function
$\tau_{SSW}(\bt)$ of the Drinfeld--Sokolov hierarchy \eqref{Tat}.

\begin{thm}\label{thm-tautauDS}
For the Drinfeld--Sokolov hierarchy \eqref{Lt2} associated to $\fg$,
the two tau functions defined by \eqref{tauDS} and by \eqref{satoDS}
satisfy
\begin{equation}\label{tautauDS}
\log\tau=\frac{\kappa}{r\,k_0}\log\tau_{SSW}.
\end{equation}
\end{thm}

In order to prove this theorem, we need the following lemma.
\begin{lem}\label{thm-resc}
Under the homogeneous gradation of $\fg$,  let $e^X$ be a well
defined element in the Kac--Moody group of $\fg$. Then for any $
Y\in\fg$, the following equality holds true
\begin{equation}\label{}
\left(e^X  Y
e^{-X}\right)_c=\frac{\kappa}{r\,k_0}\oint_{S^1}\mathrm{Tr}\left((\pd_z
e^X) Y e^{-X}\right)\frac{d z}{2\pi\mathbf{i}}.
\end{equation}
\end{lem}
\begin{prf}
In order to simplify notations, we write for any $X\in\fg$:
\[
X'=\pd_z X, \quad \la X
\ra=\oint_{S^1}\mathrm{Tr}\left(X\right)\frac{d z}{2\pi\mathbf{i}}.
\]
First of all, we have
\[
\la (\pd_z e^X) Y e^{-X} \ra= \sum_{m=1}^\infty R_m,
\]
where
\begin{align*}
R_m=\left\la \sum_{k=0}^{m-1}\frac{(-1)^k}{(m-k)! k!}(X' X^{m-k-1}+X
X' X^{m-k-2}+\dots+X^{m-k-1}X') Y X^k\right\ra.
\end{align*}
We rewrite $R_m$ to
\begin{align}\label{}
R_m=&\left\la \sum_{k=0}^{m-1}\frac{(-1)^k}{(m-k)!
k!}X'\sum_{j=0}^{m-k-1}X^j Y X^{k}\right\ra \nn\\
=&\frac{1}{m!}\left\la
X'\sum_{j=0}^{m-1}\sum_{k=0}^{m-j-1}(-1)^k\binom{m}{k}X^j Y X^{m-1-j}\right\ra \nn\\
=&\frac{1}{m!}\left\la
X'\sum_{j=0}^{m-1}(-1)^{m-1-j}\binom{m-1}{m-1-j}X^{j} Y X^{m-1-j}\right\ra \nn\\
=&\frac{1}{m!}\left\la X'(\ad_X)^{m-1} Y\right\ra, \label{Rm}
\end{align}
where in the third equality we have employed recursively the formula
\[
\binom{m-1}{k}-\binom{m}{k+1}=-\binom{m-1}{k+1}.
\]
According to the homogeneous realization of $\fg$,   we have
$N_{\rs^0}=k_0$, and it follows from \eqref{XYbr} that
\begin{equation}\label{brc}
[A(z), B(z)]_c=\frac{\kappa}{r\,k_0}\oint_{S^1}\mathrm{Tr}(\pd_z
A(z)\cdot B(z) )\frac{d z}{2 \pi\mathbf{i}}
\end{equation}
for any $A(z), B(z)\in\fg$. By using this equality and \eqref{Rm},
we obtain
\begin{align}\label{}
\la (\pd_z e^X) Y e^{-X} \ra
=&  \sum_{m=1}^\infty\frac{1}{m!}\left[X,(\ad_X)^{m-1} Y\right]_c\cdot\frac{r\,k_0}{\kappa} \nn\\
=& \sum_{m=1}^\infty\frac{1}{m!}\left(
(\ad_X)^m Y\right)_c\cdot\frac{r\,k_0}{\kappa}\nn\\
=&\frac{r\,k_0}{\kappa}(e^{\ad_X}  Y )_c \nn\\
=&\frac{r\,k_0}{\kappa}(e^X  Y e^{-X})_c.
\end{align}
The lemma is proved.
\end{prf}

\begin{prfof}{Proposition~\ref{thm-tautauDS}}
Recall  \eqref{tauDS} and by \eqref{satoDS}. For any $j\in E_{>0}$
we have
\begin{align}\label{}
\pd_{t_j}\log\tau=&-(\Ta\Ld_j\Ta^{-1})_c\nn\\
=&-\frac{\kappa}{r\,k_0}\oint_{S^1}\mathrm{Tr}\left(\pd_z \Ta
\cdot\Ld_j\Ta^{-1}\right)\frac{d z}{2\pi\mathbf{i}} \nn\\
=&\frac{\kappa}{r\,k_0}\pd_{t_j}\log\tau_{SSW},
\end{align}
where the second equality is due to Lemma~\ref{thm-resc}. Therefore
the proposition is proved.
\end{prfof}

\begin{exa} Assume the affine Kac--Moody algebra $\fg$ to be of type
$A^{(1)}_{n-1}$ with $n\ge2$. This Lie algebra contains a set of
Weyl generators as follows:
\begin{align}
&e_0=z\,e_{1,n}, \quad
 e_i=e_{i+1,i}~~~ (1\le i\le n-1), \label{eA} \\
&f_0=\frac{1}{z}e_{n,1},\quad f_i=e_{i,i+1}~~~ (1\le i\le n-1),\\
&\al^\vee_i=[ e_i,f_i] ~~~(0\le i\le n-1), \label{alvA}
\end{align}
where $e_{i,j}$ is the $n\times n$ matrix with its $(i,j)$-component
being $1$ and the others being zero. One has
\begin{equation}\label{}
\Ld=e_0+e_1+\cdots+e_n=\left(\begin{array}{ccccc}
0 & 0 & \dots & 0 & z\\
1 & 0 & \ddots &   & 0 \\
0 & 1 & \ddots & \ddots & \vdots \\
\vdots & \ddots & \ddots & 0 & 0 \\
0 &  \dots & 0 & 1 & 0
\end{array}\right).
\end{equation}
The principal Heisenberg subalgebra $\mathcal{H}$ is generated by
$\Ld_j=\Ld^j$ with $j\in E$, where $E=\Z\setminus n\Z$ is the set of
of exponents of $\fg$.

Via gauge actions, the operator $\sL$ in \eqref{sLgp} can be
converted to the following canonical form
\begin{equation}
q=q^{\mathrm{can}}=-\sum_{i=1}^{n-1}u_i \,e_{n-i,n}
\end{equation}
with scalar functions $u_i$. According to \cite{DS}, the
Drinfeld--Sokolov hierarchy \eqref{Lt2} is equivalent to the
Gelfand--Dickey (or $n$-reduced KP) hierarchy:
\begin{equation}\label{GD}
\frac{\pd L}{\pd t_j}=[(L^{j/n})_+,L],\quad j\in \Z_+\setminus
n\Z_+,
\end{equation}
where
\begin{align}
&L={\p_x}^{n}+u_1 {\p_x}^{n-2}+\dots+u_{n-2} {\p_x}+u_{n-1}, \label{LAn0} \\
&L^{1/n}={\p_x}+v_1 {\p_x}^{-1}+v_2 {\p_x}^{-2}+\cdots, \nn
\end{align}
and $(L^{j/n})_+$ means the differential part of the operator
$L^{j/n}$. Recall that the multiplication between two
pseudo-differential operators is defined by
\[
u {\p_x}^k\cdot v {\p_x}^l=\sum_{m\ge0}\binom{k}{m} u {\p_x}^m(v)
{\p_x}^{k+l-m}.
\]

For the hierarchy \eqref{GD} the tau function $\tau_{SSW}$ in
\eqref{satoDS} was also introduced in \cite{SW}, which satisfies
\begin{equation}\label{GDh}
\frac{\p^2\log\tau_{SSW}}{\p x\p t_j}=\res\,L^{j/n}, \quad j\in
\Z_+\setminus n\Z_+.
\end{equation}
Note that the residue of a pseudo-differential operator means its
coefficient of ${\p_x}^{-1}$.

On the other hand, the tau function $\tau$ of the hierarchy is
defined by \eqref{tauDS}. In this case, Theorem~\ref{thm-tautauDS}
shows
\[
\tau=\tau_{SSW},
\]
which agrees with Example~5.1 in \cite{Wu}.
\end{exa}

%\begin{exa}
%When the affine Kac--Moody algebra $\fg$ is of type
%$D^{(1)}_{n/2+1}$ with even $n$, take its matrix realization
%$\fg(A;\rs^0)=\mathfrak{so}_n[z,z^{-1}]\otimes\C c$ as in \cite{DS,
%Kac}. The coefficient on the right hand side is $\kappa/(r
%k_0)=1/2$, hence $\tau_{SSW}$ is the same as the tau function given
%in \cite{LWZ} ( see also \cite{Wu}).
%\end{exa}

Similar argument applies for  Drinfeld--Sokolov hierarchies
associated to affine Kac--Moody algebra of type $D^{(1)}_{n/2+1}$
with even $n$, see \cite{LWZ, Wu}. Furthermore, by using
Theorem~\ref{thm-tautauDS} and the appendix of \cite{Wu}, via $\tau$
one can see the relation between the tau function $\tau_{SSW}$ and
those tau functions defined by Enriquez and Frenkel \cite{EF} and by
Miramontes \cite{Mi} for Drinfeld--Sokolov hierarchies.

\begin{rmk}
In a recent work \cite{Sa} (see also \cite{BF}), Safronov proposed a
geometric definition of tau function for  Drinfeld--Sokolov
hierarchies. His tau function can be considered, in a sense, as an
algebro-geometric version of the Sato--Segal--Wilson tau function,
which was shown to coincide with the tau function given in
\eqref{tauDS}. In the present paper, we derived such an equivalence
relation in a more explicit way. We hope that results in this
direction would be helpful to understand the Virasoro symmetries for
Drinfeld--Sokolov hierarchies \cite{Wu} from the viewpoint of Kac
and Schwarz \cite{KS}, which we plan to study elsewhere.
\end{rmk}

\subsection{Tau functions of generalized Drinfeld--Sokolov hierarchies}

In the construction of generalized Drinfeld--Sokolov hierarchies by
de Groot, Hollowood and Miramnontes \cite{GHM, Mi} (c.f.
\cite{FHM}), the principal Heisenberg subalgebra $\mathcal{H}$ is
replaced by the Heisenberg subalgebra $\mathcal{H}_{[\rs']}$
corresponding to certain gradation $\rs'\in\Gm$, which is induced by
some conjugacy class of the Weyl group related to the simple Lie
algebra $\mathcal{G}$ (see \cite{Kac, KP} for details). Here we only
use the form of $\cH_{[\rs']}$ as
\[
\cH_{[\rs']}=\C c\oplus\sum_{j\in E'}\C\Ld_j,
\]
where $\Ld_j\in\fg_{j\,[\rs']}$, and $E'\equiv E'_0\mod N_{\rs'}$
with $E'_0$ being a collection of $l$ non-negative integers lower
than $N_{\rs'}$. Similar as before, the elements $\Ld_j$ are
normalized as
\begin{equation}\label{Ldjk2}
[\Ld_j, \Ld_k]=j\dt_{j,-k}c
\end{equation}
(here we avoid lengthy notations when $E'$ contains multiple degrees
$j_1, j_2, \dots, j_p$ that equal $j$, in which case
$\dt_{j_p,-k_q}$ stands for $\dt_{j,-k}\dt_{p,q}$).

Let $m=\min E'_{>0}$, and let $\Ld=\Ld_m\in\cH_{[\rs']}$ be fixed.
We only consider the case that $\Ld$ is a semisimple element
satisfying \eqref{dec2}. In this case the generalized hierarchies
are called of type I in \cite{GHM}, which contain the
Drinfeld--Sokolov hierarchies above as a particular case.

Choose an additional gradation $\rs=(s_0,s_1,\dots,s_l)\preceq\rs'$
of $\fg$, namely, $s_i\le s_i'$ for all $i$. Introduce an operator
\begin{equation}\label{}
\sL=\frac{\od}{\od x}+\Ld+q,
\end{equation}
where $q$ is a smooth function taking value in
$\left.\left(\fg_{0\,[\rs]}\cap\fg_{\le0\,[\rs']}\right)\right/\mathcal{H}_{0\,[\rs']}$.
Note that the quotient means to modulo variables along trivial or
non-independent flows of the hierarchy to be defined (see, for
example, Proposition~3.6 in \cite{GHM}).

With the gradations $\rs^0$ and $\rs^1$ in Subsection~4.1 replaced
by $\rs$ and $\rs'$ respectively, one has an analogy of
Proposition~\ref{thm-dr0}, and hence defines the generalized
Drinfeld--Sokolov hierarchy
\begin{equation}\label{Lt3}
\frac{\pd \sL}{\pd t_j}=[(e^{\ad_U}\Ld_j)_{<0\,[\rs]}, \sL], \quad
j\in E'_{>0},
\end{equation}
up to the gauge actions induced by functions in the nilpotent
subalgebra $\fg_{0\,[\rs]}\cap\fg_{<0\,[\rs']}$.

To simplify the discussions, let us take $\rs=\rs'$, hence the
hierarchy \eqref{Lt3} can be recast to a form as \eqref{Tat} of a
dressing element $\Ta$ in the Lie group of $\fg_{<0\,[\rs]}$.
Accordingly, it applies verbatim the argument in the previous
subsection of Baker function and Sato-Segal-Wilson tau function
$\tau_{SSW}$ for the Grassmannian with two groups like
\eqref{lieGm}--\eqref{lieGa}. Thus we can define $\tau_{SSW}$ to be
the tau function of the generalized Drinfeld-Sokolov hierarchy
corresponding to the Heisenberg subalgebra $\mathcal{H}_{[\rs']}$.
This definition, due to an analogy of the generalize Sato formula
\eqref{satoDS}, can be viewed as a generalization of \eqref{tauDS}
for the original Drinfeld--Sokolov hierarchies.

\begin{rmk}
The system \eqref{Lt3} with $\rs=\rs'$ is sometimes called the
generalized Drinfeld--Sokolov hierarchy of modified KdV type, which
is related to hierarchies with different $\rs$ (but the same $\rs'$)
via Miura-type transformations.
\end{rmk}

\begin{exa}
Let us take $\rs=\rs'=\rs^0$ being the homogeneous gradation. The
homogeneous Heisenberg subalgebra $\mathcal{H}_{[\rs^0]}$ contains a
basis:
\[
\set{c, \, z^j H_i \in\fg_{j h\,[\rs^0]}\mid j\in\Z, ~i=1, 2, \dots,
l}
\]
with $h=\sum_{i=0}^l k_i$ being the Coxeter number of $\fg$, and
$E'=\hbox{$l$-ple $h\Z$}$. Note that the elements $z^j H_i$ may not
satisfy the condition \eqref{Ldjk2} (linear combinations of them can
give the normalized generators $\Ld_j$ in \eqref{Ldjk2}). For the
homogeneous Drinfeld--Sokolov hierarchy, the big cell of the
Grassmannian is acted by the following subgroups of $L_{1/2}U_n$:
\begin{align}
G_-=&\left\{ e^X\in L_{1/2}U_n\mid X\in\fg_{j<0\,[\rs^0]} \right\},
\\
G_+^a=&\left\{g(\mathbf{s}; z)=\exp\left(\sum_{k\ge1; 1\le i\le
l}s_{k,i} z^k H_i\right)\in L_{1/2}U_n\mid s_{k,i}\in\R \right\}.
\end{align}
The Baker function is defined in the same way as before,
%Definition~\ref{def-bf},
and the Sato--Segal--Wilson tau function
$\tau_{SSW}$ is given by the formula \eqref{satoDS}.

From another point of view, since the affine Kac--Moody algebra
$\fg$ are realized by trace-less matrices and $H_\al$ are diagonal,
then the corresponding hierarchy can be considered as the AKNS-D
hierarchy restricted to affine Kac--Moody  algebras, see Section~9.1
in \cite{Dickey} for example (the case of type $A_1^{(1)}$).

\end{exa}

\vskip 0.5truecm \noindent{\bf Acknowledgments.} The author
C.-Z.\,W. thanks Profs. Boris Dubrovin and Youjin Zhang for helpful
discussions and constant support.

\end{document}